\RequirePackage{lineno}
\documentclass[]{aa}
\usepackage{amsmath} 
\usepackage{amssymb}

\usepackage[utf8]{inputenc}
\usepackage{epsfig}
\usepackage{url}
\usepackage{color}
\definecolor{darkgreen}{rgb}{0,.6,0}
\definecolor{linkcol}{rgb}{.6,0,0}
\usepackage[colorlinks=true,citecolor=darkgreen]{hyperref}

\usepackage{natbib}
\bibpunct{(}{)}{;}{a}{}{,}
\usepackage{subfigure}

\usepackage{upgreek}  

\def\lta{~\raise.4ex\hbox{$<$}\llap{\lower.6ex\hbox{$\sim$}}~}
\def\gta{~\raise.4ex\hbox{$>$}\llap{\lower.6ex\hbox{$\sim$}}~}

\def\msol{\ensuremath{\rm \mass_\odot}} 
\newcommand\solm\msol

\usepackage{dcolumn} \newcolumntype{d}[1]{D{.}{.}{#1}}
\usepackage{multirow}

\bibpunct{(}{)}{;}{a}{}{,}

\begin{document}

\title{Planet formation by pebble accretion in ringed disks}
\titlerunning{Planet formation in ringed disks}

\author{A. Morbidelli\inst{1}}

\authorrunning{A. Morbidelli}

\institute{
$^1$ D\'epartement Lagrange, University of Nice -- Sophia Antipolis, CNRS, Observatoire de la C\^{o}te d'Azur, Nice, France
 }

\abstract{Pebble accretion is expected to be the dominant process for the formation of massive solid planets, such as the cores of giant planets and super-Earths. So, far, this process has been studied under the assumption that dust coagulates and drifts throughout the full protoplanetary disk. However, observations show that many disks are structured in rings that {{  may be due to pressure maxima, preventing}} the global radial drift of the dust.} 
{We study how the pebble-accretion paradigm changes if the dust is confined in a ring.} 
{Our approach is mostly analytic. We derive a formula that provides an upper bound to the growth of a planet as a function of time. We also implement numerically the analytic formul\ae\ to compute the growth of a planet located in a typical ring observed in the DSHARP survey, as well as in a putative ring rescaled at 5 AU.} 
{Planet Type-I migration is stopped in a ring, but not necessarily at its center. If the entropy-driven corotation torque is desaturated, the planet is located in a region with a low density of dust, which severely limits its accretion rate. If instead the planet is near the ring's center, its accretion rate can be similar to the one it would have in a classic (ring-less) disk of equivalent dust density. However, the growth rate of the planet is limited by the diffusion of dust in the ring and the final planet's mass is bounded by the total ring's mass. The DSHARP rings are too far from the star to allow the formation of massive planets {  within the disk's lifetime}. However, a similar ring rescaled to 5 AU could lead to the formation of a planet incorporating the full {  ring}'s mass in less than 1/2 My.} 
{The existence of rings may not be an obstacle to planet formation by pebble-accretion. However, for accretion to be effective the resting position of the planet has to be relatively near the ring's center and the ring needs to be not too far from the central star. The formation of planets in rings can explain the existence of giant planets with core masses smaller than the so-called pebble isolation mass.  } 


\date{[Received / accepted]}

\maketitle
\section{Introduction}\label{sec:intro}

The formation of massive planets (cores of giant planets, super-Earths) is not yet fully elucidated. The classic model of oligarchic growth in a disk of planetesimals has difficulties producing so massive bodies within the disk's lifetime, particularly at distances typical of the giant planets of the solar system or warm jupiters around other stars (Levison et al., 2010; Morbidelli et al., 2015; Johansen and Lambrechts, 2017; {  Johansen and Bitsch, 2019}). Oligarchic growth in planetesimal disks is also particularly inefficient if the initial planetesimals are mostly $\sim 100$ km in size (Fortier et al., 2013), as predicted by the streaming instability model (Johansen et al., 2015; Simon et al., 2017) and suggested by observations of the size-frequency distribution of the remaining solar system planetesimals (for asteroids see e.g. Morbidelli et al., 2009, Delbo et al.; for Kuiper-belt objects see e.g. Morbidelli and Nesvorny, 2020).

For these reasons, a new paradigm for planet formation has been developed in the last decade, dubbed {\it pebble accretion} (Ormel and Klahr, 2010; Lambrechts and Johansen, 2012, 2014; Lambrechts et al., 2014; Levison et al. 2015; Ida et al. 2016; Ormel, 2017, to quote just a few). In this paradigm, planets grow by accreting small solid material (dust grains, pebble-sized objects) as this material drifts radially in the disk due to aerodynamic drag.  If the flux of material is sufficiently large, planetary cores of 10-20 Earth masses can form within the lifetime of the disk.

Nevertheless, the models of pebble accretion developed so far are quite simplistic because they assume a continuous flux of pebbles as if their initial reservoir had an infinite radial extension. However, protoplanetary disks are not infinitely wide. The median {  observed} sizes (in gas) of Class-I and Class-II disks are 100 and 200 AU respectively, although some exceptional ones can extend to $\sim 1,000$~AU (Najita and Bergin, 2018). {  Due to observational biases, the real medians are certainly smaller.} There is evidence that the protoplanetary disk of the solar system did not exceed 80 AU in radial extension (Kretke et al., 2012). Given the expected rates of growth and radial drift of dust ({  Brauer et al., 2008}; Birnstiel et al., 2010, 2016), in disks of these sizes the flux of pebbles is expected to last much less than the lifetime of the disk. For instance, Sato et al. (2016) estimated that the flux of pebbles should decay sharply after a time $t\sim 2\times 10^5 (r_{gas}/100 {\rm AU})^{3/2}$~y, where $r_{gas}$ is the radius of the gas-disk. As a corollary, one would expect that protoplanetary disks at several AUs from the star become dust-poor very rapidly because of dust's growth and drift (Birnstiel et al., 2010; Rosotti et al., 2019), but observations show that disks remain dust-rich (compared to gas) for at least a few My (Manara et al., 2016), given the good correlation existing between the gas accretion rate onto the central star and the disk mass in dust. There is no evidence that disks are systematically smaller in dust than gas, because the observed ratios (typically $r_{gas}/r_{dust}\sim 2$ - Najita and Bergin, 2018; Ansdell et al., 2018) could simply be the consequence of optical depth effects (Trapman et al., 2019). {  For balance, it should be said that the size of disks in gas is also very uncertain and may depend on the tracer that is used (Anderson et al., 2019).}

Another issue with the pebble accretion paradigm is that it is inefficient. To form a core of 10 Earth masses starting from a Moon-mass seed, more than 100 Earth masses of pebbles/dust have to pass through the planetary orbit, most of which are not accreted by the growing planet (Morbidelli et al., 2015; Bitsch et al., 2019). The same is true for the formation of a system of super-Earths (Lambrechts et al., 2019). These unaccreted pebbles/dust should accumulate at the inner edge of the disk, where the transition from a turbulent to a more quiescent gas is expected to create a local maximum in the pressure radial distribution (Flock et al., 2016, 2017). There, the accumulated material should form massive planets (Flock et al., 2019). The corollary is that every star should have close-in massive planets, which is not observed.

{  These problems could be solved if the dust grows much more slowly than expected in coagulation models and so drifts towards the star on longer timescale (Johansen et al., 2019). However, in this case the planets would also grow more slowly (because the pebbles are smaller). Moreover, the observations of disk at mm-wavelength suggest that mm-size dust does exist in disks. Another possibility is that} in protoplanetary disks there are impediments to dust-drift, possibly due to the presence of multiple pressure bumps (Pinilla et al., 2012). Indeed, recent high-resolution observations with ALMA show that most protoplanetary disks have ring-like structures (ALMA partnership 2015; Andrews et al., 2018; Long et al., 2018; Dullemond et al., 2018), indicating that bottlenecks for dust-drift exist at multiple locations.      

The origin of these rings is still debated. Magneto-hydrodynamical (MHD) simulations show that the surface density distribution of disks can become a radial wavy function (Bethune et al., 2017), leading to the formation of multiple pressure bumps and hence of dust rings (Riols et al., 2019). More subtle dust-gas instabilities can also generate dust rings (Tominaga et al., 2019). On the other hand, massive planets can open gaps in the gas distribution and generate pressure bumps (e.g. Zhang et al., 2018; Weber et al., 2019; Yang and Zhu, 2020; Wafflard-Fernandez and Baruteau, 2020). Thus, whether rings are a prerequisite for planet formation or a consequence of planet formation is not yet known. Similarly, for the solar system there is evidence that the dust from the outer disk did not penetrate in large quantities into the inner disk, leading to the observed dichotomy in the isotopic properties of inner and outer solar system planetesimals (Kurijer et al., 2017). Whether this separation between inner and outer disk's dust was due to the existence of a pressure bump in the disk (Brasser and Mojzsis, 2020) or to the formation of Jupiter (Kruijer et al., 2017) is under debate.

This short paper places itself in the first hypothesis (rings exist before planet formation takes place) and explores how the pebble accretion paradigm changes in presence of rings. If rings are a consequence of planet formation, what follows can be regarded as a description of how a second generation of planets can grow, from the rings induced by the formation of the first generation. {  Two important assumptions of this work are that (i) the rings are due to pressure maxima, as advocated by Dullemond et al. (2018), and (ii) they are long-lived structures which, frankly, is not known. If the rings are ephemeral, or just the result of a modulation of the dust drift speed, they probably don't have a significant impact on planet accretion.}

There are three big changes in the pebble-accretion narrative if the dust is confined in rings. First, there is no net radial drift of dust. Dust diffuses back and forth within the ring, which gives to the planet multiple chances to accrete it, alleviating the inefficiency of pebble accretion upon a single passage of a dust grain through the planet's orbit. Second, the reservoir of solid mass available for planet growth is finite, limited by the mass of the ring.  Third, there is no differential azimuthal velocity between the planet and gas/dust near the pressure maximum. Section~2 analyzes all of these aspects, leading to an analytic upper bound of the planet's growth as a function of ring mass, turbulent diffusion and dust's Stokes number. Section~3 will then present a more detailed computation of the growth of a planet in the rings observed in the DSHARP survey, discussed in Dullemond et al. (2018). The results are discussed in section~4.  

\section{Dust dynamics in a ring and its accretion by a planet}

The formation of a dust ring in the disk is most likely due to the appearance of a pressure bump in the disk, whatever the cause for this pressure bump (e.g. MHD instabilities, dust-gas instabilities, the existence of other planets). As in Dullemond et al. (2018), we assume that the radial pressure profile is Gaussian, namely:
\begin{equation}
  p(r)=p_0 \exp\left(-{{(r-r_0)^2}\over{2w^2}}\right)\ .
  \label{pressure}
\end{equation}
where $r_0$ is the center of the pressure maximum and $w$ its width.

Because of gas drag, the dust is attracted towards the pressure maximum, with a radial speed that is
\begin{equation}
  v_r(r)=-{{H^2\Omega \tau}\over{w^2(1+\tau^2)}} (r-r_0)
  \label{dustradialspeed}
\end{equation}
(see formula 44 in Dullemond et al., 2018), where $H$ is the pressure scale-height of the disk, $\Omega$ is the keplerian orbital frequency and $\tau$ is the dust's Stokes number\footnote{The original formula in Dullemond et al. uses the sound speed $c_s$, which is equal to $H\Omega$.}.

On the other hand, the dust undergoes also turbulent diffusion. The diffusion coefficient for the dust is (Youdin and Lithwick 2007):
\begin{equation}
  D_d={{D}\over{1+\tau^2}}
  \label{Dd}
\end{equation}
where $D$ is the turbulent diffusion in the gas, usually assumed to be equal to its viscosity which, following the $\alpha$-prescription of Shakura and Sunyaev (1973), gives  $D={\alpha H^2\Omega}$, where $\alpha$ is the turbulent parameter. Observations of the dust distribution in disks suggest that $\alpha=10^{-4}$--$10^{-3}$ (Pinte et al., 2016; Dullemond et al., 2018).

The balance between diffusion (which tends to disperse the dust) and the drag force (which tends to bring the dust back towards the pressure maximum) gives the dust the steady state radial distribution:
\begin{equation}
   \Sigma_d(r)=\Sigma_{d,0} \exp\left(-{{(r-r_0)^2}\over{2w_d^2}}\right)
  \label{dust}
\end{equation}
with a width $w_d$ related to the width $w$ of the pressure bump by the relationship
\begin{equation}
  w_d=w/\sqrt{1+\tau/\alpha}
  \label{wd}
\end{equation}
(Dullmond et al., 2018), which reduces to the often quoted relationship $w_d=w\sqrt{\alpha/\tau}$ for $\alpha \ll \tau$. {  Dullemond et al. (2018) show that there is typically an order of magnitude contrast between the center of the ring and its border (the ``gap'' between rings), arguing that the description given by eq. (\ref{dust}) is valid for $|r-r_0|\lesssim 2 w_d$.}

The dynamics of the growing planet responds to the surface density distribution of the gas $\Sigma_g$, which sets its migration speed and direction. The latter is related to the the pressure profile by the relationship
\begin{equation}
  p(r)=c_s^2 \Sigma_g/(\sqrt{2\pi} H)\ ,
  \label{pSigma}
\end{equation}
where $c_s=H\Omega$ is the sound speed. Assuming for simplicity $H\propto r$ (flat disk) one gets $\Sigma_g(r)\propto r^2 p(r)$, with $p(r)$ given by (\ref{pressure}). It is trivial to see that $\Sigma_g(r)$ is proportional to $r^2$ at the location of the pressure maximum $r_0$, whereas it has a maximum at $r_M=r_0 (1+\sqrt{1+8w^2/r_0^2})/2$. For $w/r_0\ll 1$ the location of the maximum of $\Sigma_g$ can be approximated by $r_M=r_0+2w^2/r_0$. Approximating $\Sigma_g(r)$ near its maximum by a parabolic profile, we then find that that $\Sigma(r)\propto 1/r$ at $r_1=r_0+3w^2/r_0$.

The planet's radial migration in disks with arbitrary surface density profiles has been studied in detail in Paardekooper et al. (2010, 2011).  They demonstrated that  the planet feels a torque $\Gamma$ from the disk that is:
\begin{equation}
  \Gamma={\Gamma_0\over{\gamma}} \left[(-2.5-1.7b+0.1a)
    +1.1f_{\rm v}(\nu,M)\left({{3}\over{2}}-a\right)\right.\nonumber
\end{equation}
\begin{equation}
   \quad  +\left. f_{\rm E}(\nu,\chi,M)\left(7.9{{\xi}\over{\gamma}}\right)\right]\ ,
  \label{torque}
\end{equation}
where $\Gamma_0=[(M/M_*) (r_p/H_p)]^2\Sigma_g(r_p) r_p^2\Omega_p^2$ is the nominal torque ($M$ and $M_*$ being the masses of the planet and the star respectively and $r_p$ the location of the planet, where the disk has height $H_p$ and rotation frequency $\Omega_p$). In (\ref{torque}), the coefficient $a$ is the exponent of the gas surface density profile, approximated as a power-law of type $\Sigma_g(r)=\Sigma_g(r_p)/(r-r_p)^a$ around $r_p$; similarly $b$ is the exponent of the disk's temperature profile, written as $T(r)=T_0(r_p)/(r-r_p)^b$.  Coherently with our assumption that the ratio $H/r$ in the disk is constant, we assume that $b=1$. The coefficient $\gamma$ in (\ref{torque}) is the adiabatic index, in general assumed to be equal to 1.4 and $\xi=b-(\gamma-1)a$. In formula (\ref{torque}) the first term within $(.)$ is due to the so-called {\it Lindblad torque}; the second one is due to the {\it vortensity-driven corotation torque} and the last is due to the {\it entropy-driven corotation torque}.

{  The functions $f_{\rm V}(\nu,M)$ and $f_{\rm E}(\nu,\chi,M)$ are between 0 and 1 and depend on the disk's viscosity $\nu$ and thermal conductivity $\chi$ and the planet's mass (Paardekooper et al., 2011). In particular $f_{\rm V}(\nu,M)\to 0$ as $\nu\to 0$; in other words, the vortensity-driven corotation torque vanishes in the inviscid limit, which is called {\it torque saturation}. In this case, the planet is capable of changing the surface density profile of the disk in its corotation region towards the slope with $a=3/2$, which erases the corotation torque. However, this is true only under the hypothesis that the viscous torque is responsible of the original density profile in the disk. But, in a low-viscosity disk, there must be other processes that are responsible for shaping the pressure maximum, for instance the magnetic stress (Bethune et al., 2017), or the torque from an already existing planet opening a gap between two adjacent rings. It is reasonable to expect that these processes would restore the original disk's profile in the planet's corotation region, fighting against the planet's action. Thus, we may assume that $f_{\rm V}(\nu,M)\sim 1$ in our case, although a specific evaluation would be needed once the actual origin of the pressure maximum is understood.}

{  The function $f_{\rm E}(\nu,\chi,M)\to 0$ if $\chi\to 0$, i.e. in the {\it adiabatic limit}, where thermal diffusivity is null. Instead, in the {\it isothermal limit} ($\chi\to\infty$) the function $f_{\rm E}$ tends to a limit value $f_{\rm E}^\infty (\nu,M)<1$, which also tends to zero if $\nu\to 0$. Thus a variety of values of $f_{\rm E}$ can be envisioned.}

\begin{figure*}[t!]
\centerline{\includegraphics[width=18.cm]{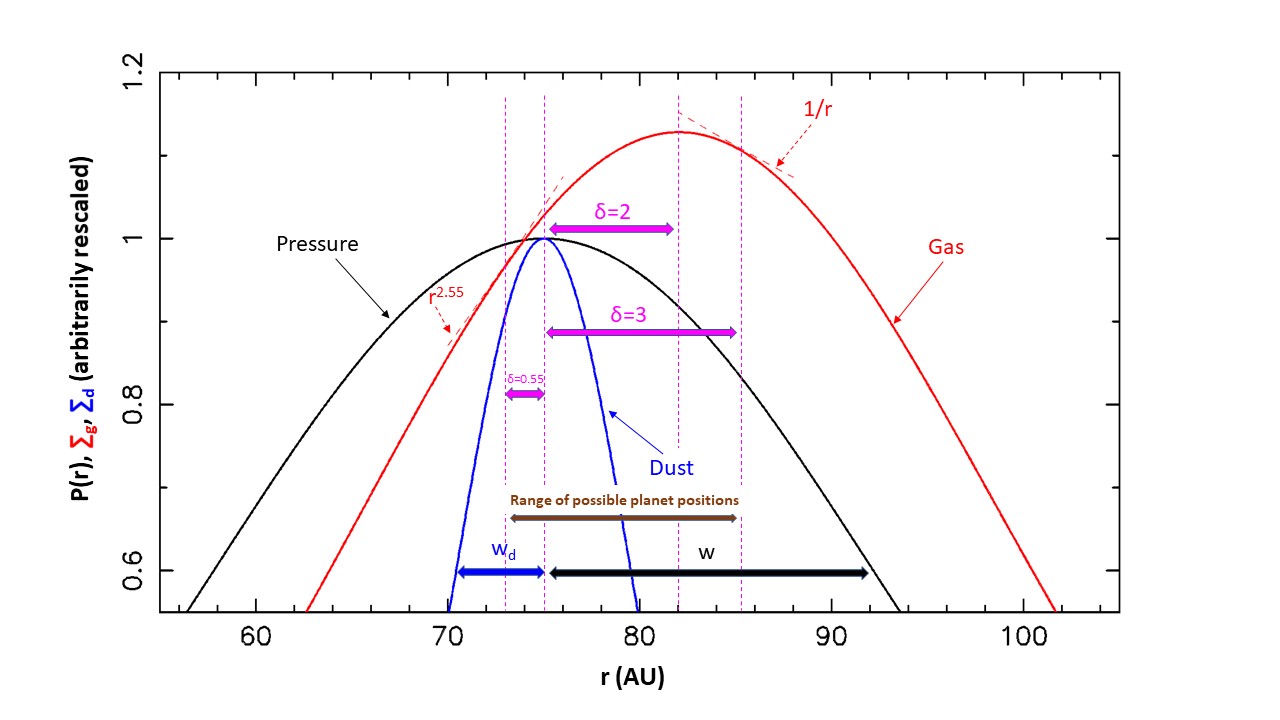}}
\caption{\small The radial profiles of pressure (black), gas surface density (red) and dust surface density (blue) for the DHSARP ring B77, from Dullemond et al. (2018) - see Sect. 3. The vertical scale of each profile has been renormalized for illustrative purposes. The dashed thin red curves show the profiles $r^{2.55}$ and $1/r$, and their tangential points to the red solid curve shows the locations where the gas surface density profile has locally these profiles. These two locations delimit the range of possible positions of the planet at the migration trap, depending on the degree of saturation of the vortensity-driven corotation torque. Their distances relative to the location of the pressure maximum are given in terms of the parameter $\delta$, as well as the location of the maximum of the gas surface density distribution. The widths $w$ and $w_d$ in (\ref{pressure}) and (\ref{dust}) are also graphically represented.}
\label{densities}
\end{figure*}

There is a planet trap -where the planet stops migrating (Masset et al., 2006b)-
where the term within $[.]$ in (\ref{torque}) vanishes. {  Assuming the limit case $f_{\rm V}=f_{\rm E}=1$, this happens at the location where $a\sim 1$ (precisely: 0.95). i.e. $\sim 3w^2/r_0$  beyond the pressure bump. If instead $f_{\rm V}=1$ but $f_{\rm E}=0$ the solution of the $[.]=0$ equation occurs for $a=-2.55$, i.e. $\sim 0.5w^2/r_0$ inwards of the pressure bump (which corresponds to $a=-2$). Notice that, if also the vortensity-driven torque were strongly saturated ($f_{\rm V}\sim 0$), the equation $[.]=0$ would not have solution: this means that there would be no planet trap and the planet would migrate out of the ring. Because this study makes sense only if the planet is confined in the ring, we assume that the vortensity-driven corotation torque is strongly desaturated ($f_{\rm V}\sim 1$, as justified above), so that the planet trap is} at a distance $\delta w^2/r_0$ from the pressure bump, with $\delta$ between $-0.5$ and $3$ depending on the actual degree of saturation of the entropy-driven corotation torque. {  Thus, $\delta$ is a measure of where the planet trap is relative to the pressure maximum, in units of $w^2/r_0$.} The radial profiles of pressure, gas and dust surface densities as well as the range of possible locations of the planet trap and corresponding values of $\delta$ are illustrated in Fig.~\ref{densities}. {  For completeness, had we assumed a {\it flared} disk, i.e. with $H/r\propto r^{2/7}$ and $b=3/7$, the planet trap would be located at $\delta=$0.15--2 (again, depending on the saturation of the entropy-driven torque) and the maximum of the gas surface density distribution would be at $\delta=12/7$.}

The planet accretes pebbles/dust at a rate
\begin{equation}
  \dot{M}=C\Sigma_d(r_p)
  \label{pa}
\end{equation}
where $\Sigma_d(r_p)$ is the density of dust at the planet's location and the coefficient $C$ depends on the accretion cross-section of the planet and the velocity of the dust relative to the planet, according to the various accretion regimes (Bondi accretion, Hill accretion, 2D, 3D,...; see the compendium in Ida et al., 2016).

From (\ref{dust}) the density of the dust at the planet trap is
\begin{equation}
  \Sigma_d(r_p)=\Sigma_{d,0} \exp\left(-{{\delta^2w^4}\over{2r_0^2w_d^2}}\right)=
  \Sigma_{d,0} \exp\left(-{{\delta^2w^2(\alpha+\tau)}\over{2\alpha r_0^2}}\right)
  \label{dust-trap}
\end{equation} 
where we have expressed $r_p-r_0$ in terms of $\delta$ and used (\ref{wd}) to pass from the second to the third expression. Notice that, in a low viscosity disk with $\alpha \ll \delta^2w^2\tau/(2r_0^2)$, the dust density (\ref{dust-trap}) is exponentially small {  at the location $\delta$ of the planet}, so that the planet can accrete only at a slow rate. This is a first big difference with respect to a disk without pressure traps, where the dust density scales smoothly with $r$ (approximately $\sim 1/\sqrt{r}$, assuming $\tau$ and the dust mass-flux to be independent of $r$), so that, wherever the planet is, it can accrete at a comparable rate.

If instead $\delta$ is small, $\Sigma_d(\delta)\sim \Sigma_{d,0}$. The velocity of the gas relative to the planet becomes small, being
\begin{equation}
  \Delta v= {1\over 2} {{H^2\Omega}\over{r}} {{{\rm d}\log P}\over{{\rm d}\log r}} = {{1}\over{2}} {{\delta H^2\Omega}\over{r_0}}\ .
    \label{delta-v}
\end{equation}
However, the planet's accretion rate in the 3D Bondi regime, for a given particle's $\tau$,  is independent of $\Delta v$ and it is equal to $3/2 \pi R_H^3 \Omega \rho_d$, where  $R_H =r_p[(M/(3M_*)]^{1/3}$ is the Hill radius and $\rho_d$ is the volume density of dust near the midplane (Ida et al., 2016; Ormel, 2017). Thus the accretion rate is the same as in a disk without pressure bump and dust surface density $\Sigma_{d,0}$.

A caveat, however, needs to be mentioned. At $\delta\sim 0$ the gas near the planet's orbit is not pressure-supported and thus it has to follow horseshoe streamlines relative to the planet even if the planet's mass is small. The half-width of the horseshoe region is $x_s=1.16r_p (M/M_*)^{1/2}(H/r)^{-1/2}$ when the planet mass is small and tends to $\sim 2.5 R_H$ when $M/M_*\gtrsim 10^{-4}$ (Masset et al., 2006a). If one neglects the perturbations of the planet on the gas, the nominal accretion radius in the Hill regime is $r_{acc}=(\tau/0.1)^{1/3} R_H$. If $x_s\ll r_{acc}$, the dust reaches the Hill radius of the planet without being substantially affected by the distortions of the gas streamlines, so the nominal accretion radius is the correct radius. But if $x_s\gtrsim r_{acc}$ the specific flow of the gas becomes important. Only the dust carried along a narrow stream of gas around the separatrix of the coorbital motion is accreted by the planet. The radial width $W(M,\tau)$ of this stream is a complicated function of planet mass and particles' Stokes number and is described in Kuwahara and Kurokawa (2020). On the other hand, the shear velocity of this dust relative to the planet is $3/2 x_s \Omega$, which is larger than $3/2 r_{acc}\Omega$. Using numerical simulations, Kuwahara and Kurokawa showed that, if the dust density is radially uniform, these two effects approximately cancel out and the actual accretion rate onto the planet is almost the same as that computed, a priori incorrectly, using the nominal accretion radius. However, in the case of a ring, the density of dust (\ref{dust}) is not uniform. Thus, the estimate of the dust accretion rate should be decreased by a factor $\exp(-x_s^2/2w_d^2)$. In most cases this factor will be $\sim 1$, but it can be significantly smaller if the ring is narrow (small $w_d$) and the planet massive (large $x_s$). 

In any case, the planet cannot accrete more dust that the ring can deliver by diffusion and/or advection. Because the dust density profile (\ref{dust}) is set by the equilibrium between diffusion and advection, we can use either of them to evaluate the dust delivery rate to the planet's orbit. Using the advection term (\ref{dustradialspeed}) and expression (\ref{dust-trap}), we find that the planet's accretion rate is capped by (assuming $\alpha\ll\tau\ll 1$):
\begin{equation}
  \dot{M}_{Max}=2 \times 2\pi r_p \Sigma_d(r_p) v_r(r_p) = A \tau\delta\Sigma_{d,0} \exp\left(-{{\delta^2w^2\tau}\over{2\alpha r_0^2}}\right)\ ,
    \label{cap}
\end{equation}
with $A=4\pi r_p H_p^2\Omega_p/r_0$.  Notice the first factor of 2 at the right hand side of the first equal sign, which is due to the fact that dust is delivered from both sides of the planet's orbit due to diffusion, unlike in a disk dominated by advection, where the material flows only from one side. For the reasons explained above if $\delta\sim 0$ the dust flux should be computed by setting $\delta=max(r_{acc},x_s) r_0/w_d^2$. For comparison, the maximal accretion rate in a disk without pressure bump and dust density $\Sigma_{d,0}$ would be $\dot{M}_{Max}=4\pi r_p^2\Omega_p \tau \eta \Sigma_{d,0}$, i.e. a factor $\sim 1/\delta \exp[\delta^2w^2\tau/(2\alpha r_0^2)]$ larger than the one in (\ref{cap}). Formula (\ref{cap}) has a maximum for $\delta=\sqrt{\alpha/\tau} r_0/w$, where $\dot{M}_{Max}=A\sqrt{\alpha\tau}r_0/w\Sigma_{d,0}\exp(-1/2)$, i.e. is proportional to the square root of the diffusion coefficient in the disk.

Another aspect to consider about pebble accretion in a ring is that the reservoir of solid mass is finite and equal to the dust mass of the ring. As the planet is growing by accreting dust, the total dust mass in the ring decreases as $M_{\rm ring}=M_{\rm ring}^{\rm init} -M$, where  $M_{\rm ring}^{\rm init}$ is the initial dust mass in the ring and $M$ is the mass of the planet. To obtain an upper bound on the planetary mass as a function with time, we assume that the planet always accretes at the maximal possible rate (\ref{cap}). By approximating
\begin{equation}
  \Sigma_{d,0}={{M_{\rm ring}}\over{(2\pi)^{3/2}r_0 w_d}}={{M_{\rm ring}}\over{(2\pi)^{3/2}r_0 w}}\sqrt{{{\tau}\over{\alpha}}}\ ,
\end{equation}
we find that the planet's mass would evolve according to the differential equation:
\begin{equation}
  \dot{M}={{A}\over{(2\pi)^{3/2}r_0 w}} \sqrt{{{\tau^3}\over{\alpha}}} (M_{\rm ring}^{\rm init}-M)\delta\exp\left(-{{\delta^2w^2\tau}\over{2\alpha r_0^2}}\right)
  \label{dotMmax}
\end{equation}
whose solution is:
$M(t)=M_{\rm ring}^{\rm init}[1-\exp(Bt)]$,
with
\begin{equation}
  B={{A}\over{(2\pi)^{3/2}r_0 w}} \sqrt{{{\tau^3}\over{\alpha}}}\delta\exp\left(-{{\delta^2w^2\tau}\over{2\alpha r_0^2}}\right)\ .
\end{equation}
Thus, initially the planet grows with a rate that can be small if $\alpha r_0^2/\tau w^2 \ll 1$ or $\tau^3/\alpha \delta^2 \ll 1$ (remember it's an upper bound); then the rate slows down as the mass of the planet approaches asymptotically the limit $M_{\rm ring}^{\rm init}$.

\section{An application: growth of a planet in a DSHARP ring}

To make a quantitative calculation of planet accretion in a dust ring, we consider the rings studied in Dullemond et al. (2018). In particular we choose the ring denoted B77 in that paper, as it is the ``median ring'' in terms of distance from the star and width. The properties of this ring given in Dullemond et al. are: $r_0=75$~AU, $T=22$~K,  $w_d=4.5$~AU, $M_{\rm dust}=40M_\oplus$. These properties are deduced directly from the observations. The maximal width of the pressure bump $w$ is estimated from the distance between adjacent rings and is $w=17$~AU. The radio $w_d/w$ sets $\alpha/\tau=7.7\times 10^{-2}$. Educated guesses on the gas density ($\Sigma_g=10$g/cm$^2$), and dust size then lead to $\tau=3\times 10^{-3}$ and $\alpha=2.3\times 10^{-4}$. Even if there are large uncertainties on some of these parameters, we are going to adopt them as fiducial numbers, for sake of example. We also assume for simplicity that the central star has the mass of the Sun, even though the real star is $0.8 M_\odot$. These parameters have been used also to draw Fig.~\ref{densities}, for illustrative purposes. 

We consider a planetary seed with $M=0.1M_\oplus$. {  We first assume for sake of example that the planet is fixed at three different locations ($\delta=3, -0.5$ or $0$); then we follow the expected location of the planet as its mass grows, using the Paardekooper et al. (2011) formul\ae\ for torque saturation.}

For the computation of the coefficient $C$ in (\ref{pa}) we compute the fraction of the local dust density accreted according to the Bondi, Hill, 2D or 3D accretion regimes, following the formul\ae\ in Ida et al. (2016).

\begin{figure}[t!]
\centerline{\includegraphics[width=9.5cm]{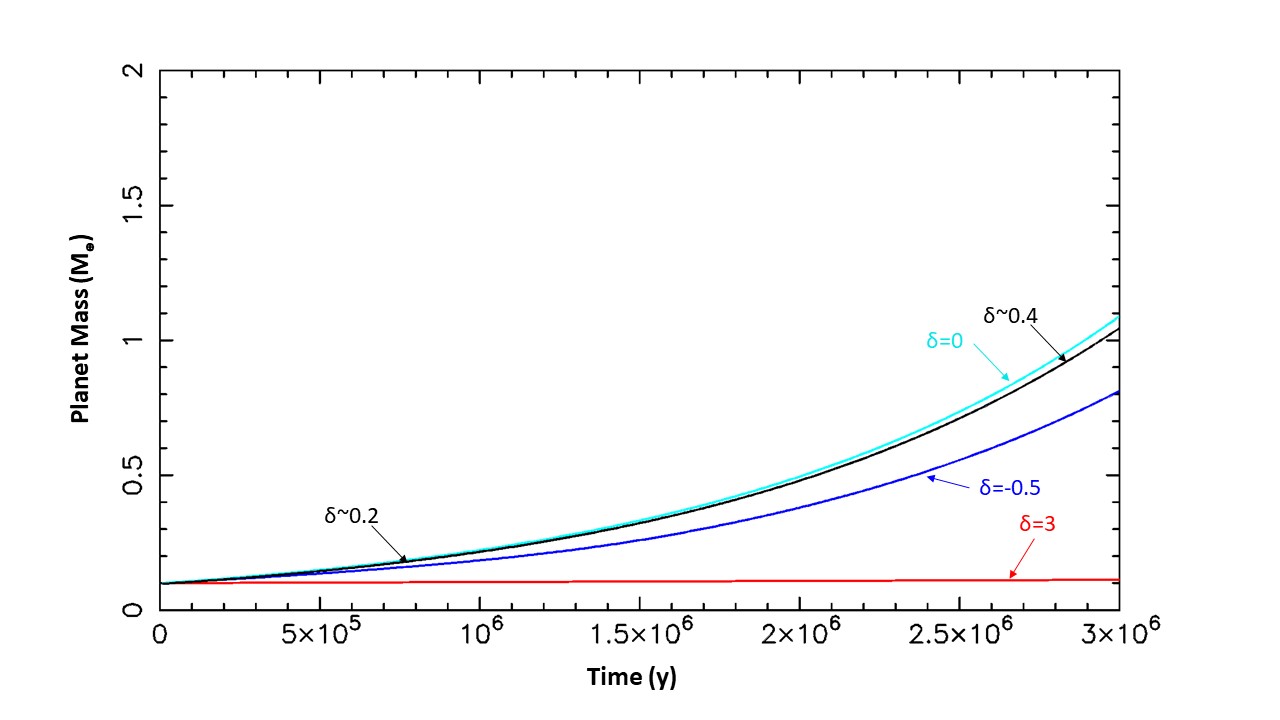}}
\caption{\small The evolution of the mass of a planet of initially 0.1~$M_\oplus$ accreting pebbles in the B77 DHARP ring, for three different locations (i.e. values of $\delta$) relative to the position of the pressure maximum {  as well as a case (black curve) where the planet follows the (mass-dependent) location of the planet trap as it grows, and therefore $\delta$ changes with time (from $\sim 0.2$ to $\sim 0.4$).}}
\label{growth}
\end{figure}

If the planet is at $\delta=3$, we find that the planet starts (and remains) in the 3D Bondi accretion regime, so that
\begin{equation}
  C= {{\sqrt{8\pi}}\over{H_d}} R_B^2 {{t_s}\over{t_B}} \Delta v
\end{equation}
where $\Delta v$ is given in (\ref{delta-v}), the Bondi radius is $R_B=GM/(\Delta v)^2$, the crossing time is $t_B=R_B/\Delta v$ and the stopping time is $t_s=\tau/\Omega$. The mass evolution of the planet is shown over the first 3 My in Fig.~\ref{growth} (red curve). We have kept the disk properties fixed over this amount of time, for simplicity, but it would not be realistic to continue beyond this timescale, because the disk's gas and dust density should decay considerably. Also, we have no knowledge of the actual lifetime of dust rings. The planet's mass seems constant, because it increases by only 12\% on the considered timescale, which is invisible at the scale of the plot. This is because at $\delta=3$ the dust density is very reduced, {  as it can be seen by extrapolating the blue curve below the lower boundary of} Fig.~\ref{densities}.

If instead the planet is at $\delta=-0.5$, the dust density is higher and the growth is significantly faster. After 1.6~My the planet has reached $0.27 M_\oplus$ and the accretion switches to the 3D Hill regime, so that
\begin{equation}
  C= {{3\sqrt{\pi}}\over{\sqrt{8}H_d}} R_H^3 {{\tau}\over{0.1}} \Omega\ .
\end{equation}
The planet reaches a final mass of 0.8~$M_\oplus$ (Fig.~\ref{growth}, blue curve).

If the planet at the dust trap it can grow at the maximal speed. In this case, the planet reaches $1.1 M_\oplus$ in the end (Fig.~\ref{growth}, cyan curve), with a growth that is basically identical to that of a planet in a disk with dust density $\Sigma_{d,0}$ without pressure bumps, because $x_s\ll 2\sqrt{2} w_d$ throughout the planet's growth. In all three cases we considered, the growth rate is much smaller than the maximum rate reported in (\ref{dotMmax}).

{  To add a bit of realism, we now compute, for each planet mass, the expected location of the planet's trap accounting for the saturation of the corotation torques, and compute the accretion rate of the planet at that location. To evaluate the saturation of the corotation torques, we use the formul\ae\ (52) and (53) in Paardekooper et al. (2011) using the disk parameters reported above, and an opacity $\kappa=0.03$~cm$^2$/g. The result is illustrated by the black curve, which is very close to the cyan curve for the planet at the pressure maximum. Indeed, in this case the planet starts at $\delta=0.2$, i.e. very close to the pressure maximum, and slowly moves away from it, reaching $\delta=0.4$ at the end of the simulation, as labeled on the plot. Throughout the simulation, the saturation functions $f_{\rm V}$ and $f_{\rm E}$ are very close to the values corresponding to the linear components of the torques, namely 0.7/1.1 and 1.7/7.9 respectively (Paardekooper et al., 2011). In fact, for the considered planetary masses the ratios $\nu r_p/(x_s^3 \Omega)$ and $\chi r_p/(x_s^3 \Omega)$, which enter in the saturation functions, are both much larger than unity, so that the system is in the isothermal, high-viscosity regime (Paardekooper et al., 2011).}  

We conclude that planetary seeds are unlikely to become a sizable (observable) planet in this ring (and probably all rings) observed by the DSHARP survey. This is consistent with the fact that we observe these rings. In fact, if the rings could form a planet containing most of their mass within the lifetime of the disk, they would be no longer visible.  

It is therefore interesting to rescale the considered ring into the inner disk. For simplicity, we multiply $r_0, w$ and $w_d$ by 0.067, so that $r_0=5$~AU. We keep the same values of $\alpha$ and $\tau$. We multiply the mass of the ring by $\sqrt{0.067}$, which is equivalent to assuming a disk's global surface density profile $\propto 1/r^{3/2}$.  {  We also reset the temperature to $T=120$~K and the aspect ratio to $H/r=0.05$, which are appropriate for a disk at 5 AU.}

\begin{figure}[t!]
\centerline{\includegraphics[width=9.5cm]{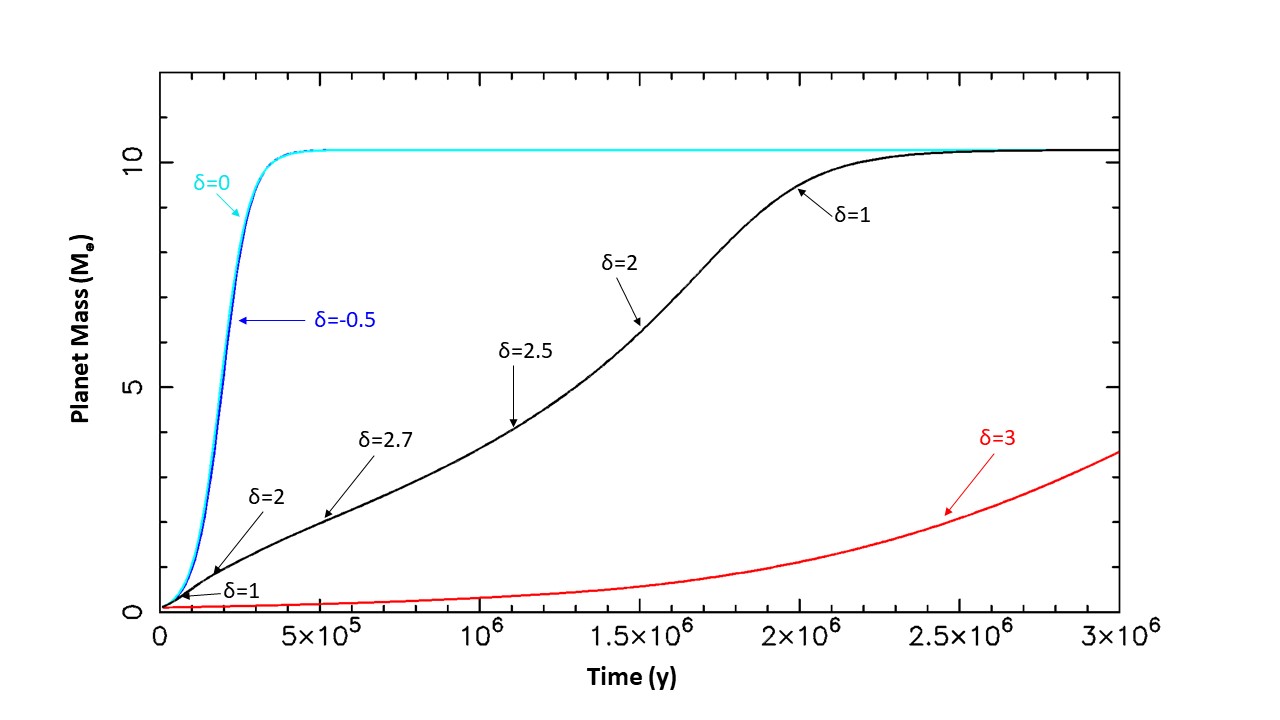}}
\caption{\small The same as Fig.~\ref{growth}, but rescaling the geometry of the B77 DHSARP ring at 5 AU and increasing the dust surface density $\propto 1/r^{3/2}$.}
\label{growthInner}
\end{figure}

The mass growths for planets situated at $\delta=3, -0.5$ and $0$ in such rescaled ring are illustrated in Fig.~\ref{growthInner}. Both the planets located at $\delta=0$ and $-0.5$ {  grow at almost indistinguishable rates} and show the saturation effect discussed at the end of Sect.~2. In fact, they accrete all the available mass ($10.2 M_\oplus$) in less than 1/2~My. {  The planet at $\delta=3$ grows much more slowly, although faster than in the ring at 75~AU, and reaches in this case $3.5 M_\oplus$ at the end of the simulation. The planet that is allowed to migrate as it grows (black curve) shows a very interesting behavior. Initially it is at $\delta=0.6$, but the planet trap moves farther away from the pressure maximum as the planet grows, reaching $\delta=2.7$ at $M=3 M_\oplus$. This limits the planet's accretion rate relatively to the cases with $\delta=0.5$ or 0 showed before,  because the density of dust at large $\delta$ is lower. Fig.~\ref{growthInner} labels the value of $\delta$ as a function of the planet mass. For larger masses, the planet migrates back towards the pressure maximum. This behavior is the effect of the corotation torques;  $f_{\rm V}$ and $f_{\rm E}$ are initially close to the values corresponding to the linear components of the torques, then they evolve towards $\sim 1$ as the planet grows and furthermore they decrease as the ratios $\nu r_p/(x_s^3 \Omega)$ and $\chi r_p/(x_s^3 \Omega)$ approach unity. As a result of this back and forth migration relative to the center of the ring, the planet reaches the maximal mass of $10.2 M_\oplus$ at 2.5~My. However, we warn that the saturation functions $f_{\rm V}$ and $f_{\rm E}$ are very sensitive on the disk's parameters, and different assumptions may lead to different results and evolutions. Here we assumed an opacity $\kappa=3$~cm$^2$/g. A lower values of $\kappa \Sigma_g$ would have kept the planet closer to the pressure maximum, enhancing its accretion rate. Also, remember that it is unclear how saturation would occur in presence of additional forces (e.g. magnetic stresses) shaping the disk's density profile. } 

Of course, we do not know whether rings of this kind exist in the inner part of protoplanetary disks, but if they do they {  can be} effective sites to produce giant planet cores, {  provided that the planet remains in the proximity of the pressure maximum.}

\section{Conclusions}

Pierre-Simon de Laplace is recognized to have been the first scientist to theorize -back in 1796- the formation of a circumstellar disk due to angular momentum conservation during the contraction of gas towards the central star (Emmanuel Kant also had envisioned the formation of a disk, but his reasoning was not scientifically correct). But less known is that Laplace, to explain the final formation of a discrete set of planets, envisioned that the circumstellar disks would break in rings,  each ring giving origin to one planet. Astonishingly, modern resolved observations of protoplanetary disks (Andrews et al., 2018) indeed show that most of them are structured in rings.

In this paper we have examined how the now popular paradigm of planet formation by pebble accretion changes if the rings are due to pressure maxima that prevent the global drift of dust in the disk. We have discussed in the Introduction that the radial drift of dust through the entire disk would lead to consequences (in terms of dust distribution, dust/gas ratio, formation of planets at the inner disk's edge) that are inconsistent with the observations. This suggests that pressure maxima are ubiquitous in protoplanetary disks (Pinilla et al., 2012), probably also in the inner parts not yet observationally resolved by current instruments.

In the analytic part of this paper (Section 2), we have pointed out that planet Type-I migration is probably stalled in a ring, but the location at which the planet is at rest is not necessarily the pressure maximum. It depends on the degree of saturation of the so-called {\it corotation torques}, particularly the one depending on the entropy gradient (Paardekooper et al., 2010, 2011). Thus, the dust accumulation may be offset from the planet's radial position by a significant amount. In this case the planet finds itself in a low-density region of dust and its accretion is necessarily slow.

If instead the planet location is near the pressure maximum, the dust density is high, but pebble accretion still proceeds differently than in the classic case where dust has a global radial drift. First, the difference in orbital velocity between the planet and the gas/particles is much smaller (it is null at the exact pressure maximum). Thus the Bondi radius becomes very large and the accretion regime transits to the Hill regime already at small planet's masses (Ormel, 2017). Moreover, in absence of differential azimuthal velocity imposed by the gas pressure gradient, the motion of dust relative to the planet is governed by 3-body coorbital dynamics instead of being a simple flyby+drag process (Kuwahara and Kurokawa, 2020). We showed, however, that both these aspects have moderate effects on the accretion rate of the planet, which remains similar to the one that the same planet would have in a disk without pressure maximum but with the same dust surface density.

More relevant is the fact that the planet cannot accrete more mass than that advected towards the planet's orbit in the ring. The density profile in the ring is the result of the balance between the drift of dust towards the pressure maximum and the turbulent diffusion in the disk (Dullemond et al., 2018). Far from the pressure maximum the drift is fast but the equilibrium surface density of dust is exponentially low; approaching the pressure maximum it's the drift speed that tends to zero. Thus, the maximal accretion rate is in general significantly smaller than the maximal accretion rate in a classic, power-law disk and it can, in some cases, effectively taper the mass growth of the planet. Moreover, the reservoir of mass available to the planet is finite (i.e. the total mass of the ring) so that the maximal planet's mass is bounded. We have obtained an analytic formula of the mass evolution of a planet assuming that it accretes at the maximal rate until exhausting the ring's dust reservoir. Of course this formula is derived from very optimistic assumptions and serves just as an upper bound of the real mass evolution.   

In Section 3 we have computed the accretion rate of a planet in a typical ring observed by the DSHARP survey (Dullemond et al., 2018), using a numerical implementation of the analytic formul\ae. We have shown that if the planet feels unsaturated corotation torques it basically does not grow because it lays too far from the peak of the ring's dust density. If the planet lays closer to the ring' s center its accretion is more significant, but nevertheless insufficient to grow the planet beyond one Earth masses in 3 My.  But it is fair to say that this inefficiency of planet formation is not due to the dust being confined in a ring, but to the long orbital timescales and low dust densities, even at peak values.

This result suggests that it is unlikely that the DSHARP rings themselves are the result of the formation of a first generation of planets. In fact, in a disk with uniform density $\Sigma_{d,0}$, the formation of the first generation of planets would still have taken as long as for the planet illustrated by the magenta curve in Fig.~\ref{growth}, reaching at most $\sim 1 M_\oplus$. And a planet with this mass is not enough to form a pressure bump in a disk with 10\% aspect ratio (Crida et al., 2005). {  Only planets growing when the disk had a higher dust density could have become more massive (Manara et al., 2018). However,} remember that without pre-existing pressure bumps the dust would have drifted towards the star on a timescale of 5,000 orbital periods (assuming a differential velocity of the gas with respect to the Keplerian velocity of $0.1 c_s$ and $\tau=3\times 10^{-3}$). Thus, the first generation of planets would have rapidly stalled growing due to the absence of solid material. These conclusions are in apparent conflict with Pinte et al. (2020), who claim the detection of  velocity ``kinks'' in 8 of 18 circumstellar disks observed by the DSHARP program, which could be due to the presence of planets. However, we note that, if confirmed, these kinks would require the existence of planets of multiple Jupiter masses; massive planets at such large distances could have formed by gravitational instability (see e.g. Kratter and Lodato, 2016 for a recent review) rather than core-accretion. 

We have then ``rescaled'' the considered ring to 5~AU, reducing its widths and vertical thickness proportionally to its radial location and scaling the dust surface density $\propto 1/r^{3/2}$. In this case, if the planet is close to the ring's center, it can accrete the whole mass of the ring within half My and can become a giant planet core. {  A planet let free to position itself within the ring according to the mass-dependent partial saturation of its corotation torques can also accrete the full mass of the ring, although on a longer timescale.}

This experiment reveals another difference with respect to the pebble-accretion paradigm in absence of pressure bumps. In that case, the growth of a planet continues until it reaches the so-called {\it pebble-isolation mass} (Lambrechts et al., 2014; Bitsch et al., 2018), at which it creates a pressure bump that stops the flux of pebbles. When this happens there is no additional energy deposited by the accretion of solids, the core can cool off and it can start accreting gas (Lambrecths et al., 2014). This predicts that the cores of giant planets should always be of the order of the pebble isolation mass. For a planet {  in a disk with $H/r=0.05$ (a typical value at 5~AU for a viscous accretion of $\sim 10^{-7} M_\odot$/y; Bitsch et al., 2014)}, this should be $\sim 20 M_\oplus$. But the core of Jupiter could be smaller than this estimate by a factor of $\sim 3$ (Wahl et al., 2017); there could even be no compact core (Debras and Chabrier, 2019). {  A pebble-isolation mass of $\sim 5$--7 $M_\oplus$ would require $H/r\lesssim 0.03$ which, at 5~AU, implies a very late disk with no accretional heating; but in this case it would be hard to accrete the massive envelope around the core.}    A suggested solution of this conundrum is that part of the core has been eroded into the envelope (Stevenson, 1985; Guillot et al., 2004). However, formation in a ring could be another elegant solution. In fact, accretion stops when the mass of the ring is exhausted which can happen well before the pebble isolation mass is reached. From that point, accretion of gas could start. This would produce giant planets with small cores.

\section{Acknowledgments}
I wish to thank K. Batygin, B. Bitsch, A. Crida, A. Johanen, M. Lambrechts and E. Lega for their comments that have improved this manuscript. This work is part of the project Gepard, funded by the French ANR and German DFG, focused on migration in low-viscosity disks.  

\section{References}

\begin{itemize}

\item ALMA Partnership, Brogan, C.~L., P{\'e}rez, L.~M., et al.\ 2015, \apjl, 808, L3
\item Anderson, D.~E., Blake, G.~A., Bergin, E.~A., et al.\ 2019, \apj, 881, 127
\item Andrews, S.~M., Huang, J., P{\'e}rez, L.~M., et al.\ 2018, \apjl, 869, L41
\item Ansdell, M., Williams, J.~P., Trapman, L., et al.\ 2018, \apj, 859, 21
  \item B{\'e}thune, W., Lesur, G., \& Ferreira, J.\ 2017, \aap, 600, A75
\item Birnstiel, T., Dullemond, C.~P., \& Brauer, F.\ 2010, \aap, 513, A79
\item Birnstiel, T., Fang, M., \& Johansen, A.\ 2016, \ssr, 205, 41
\item Bitsch, B., Morbidelli, A., Lega, E., et al.\ 2014, \aap, 564, A135
  \item Bitsch, B., Morbidelli, A., Johansen, A., et al.\ 2018, \aap, 612, A30
\item Bitsch, B., Izidoro, A., Johansen, A., et al.\ 2019, \aap, 623, A88
\item Brasser, R., \& Mojzsis, S.~J.\ 2020, Nature Astronomy, 8
  \item Brauer, F., Dullemond, C.~P., \& Henning, T.\ 2008, \aap, 480, 859
  \item Crida, A., Morbidelli, A., \& Masset, F.\ 2006, \icarus, 181, 587
  \item Delbo', M., Walsh, K., Bolin, B., et al.\ 2017, Science, 357, 1026
    \item Debras, F., \& Chabrier, G.\ 2019, \apj, 872, 100
    \item Dullemond, C.~P., Birnstiel, T., Huang, J., et al.\ 2018, \apjl, 869, L46
  \item Flock, M., Fromang, S., Turner, N.~J., et al.\ 2016, \apj, 827, 144
  \item Flock, M., Fromang, S., Turner, N.~J., et al.\ 2017, \apj, 835, 230
    \item Flock, M., Turner, N.~J., Mulders, G.~D., et al.\ 2019, \aap, 630, A147
    \item Fortier, A., Alibert, Y., Carron, F., et al.\ 2013, \aap, 549, A44
      \item Guillot, T., Stevenson, D.~J., Hubbard, W.~B., et al.\ 2004, Jupiter. The Planet, Satellites and Magnetosphere, 35
  \item Ida, S., Guillot, T., \& Morbidelli, A.\ 2016, \aap, 591, A72
\item Johansen, A., Mac Low, M.-M., Lacerda, P., et al.\ 2015, Science Advances, 1, 1500109
\item Johansen, A., \& Lambrechts, M.\ 2017, Annual Review of Earth and Planetary Sciences, 45, 359
\item Johansen, A., \& Bitsch, B.\ 2019, \aap, 631, A70
  \item Johansen, A., Ida, S., \& Brasser, R.\ 2019, \aap, 622, A202
  \item Kratter, K., \& Lodato, G.\ 2016, \araa, 54, 271
\item Kretke, K.~A., Levison, H.~F., Buie, M.~W., et al.\ 2012, \aj, 143, 91
  \item Kruijer, T.~S., Burkhardt, C., Budde, G., et al.\ 2017, Proceedings of the National Academy of Science, 114, 6712
\item Kuwahara, A., \& Kurokawa, H.\ 2020, \aap, 633, A81
\item Lambrechts, M., \& Johansen, A.\ 2012, \aap, 544, A32
\item Lambrechts, M., \& Johansen, A.\ 2014, \aap, 572, A107
  \item Lambrechts, M., Morbidelli, A., Jacobson, S.~A., et al.\ 2019, \aap, 627, A83
  \item Lambrechts, M., Johansen, A., \& Morbidelli, A.\ 2014, \aap, 572, A35
  \item Levison, H.~F., Thommes, E., \& Duncan, M.~J.\ 2010, \aj, 139, 1297
  \item Levison, H.~F., Kretke, K.~A., \& Duncan, M.~J.\ 2015, \nat, 524, 322
    \item Long, F., Pinilla, P., Herczeg, G.~J., et al.\ 2018, \apj, 869, 17
    \item Manara, C.~F., Rosotti, G., Testi, L., et al.\ 2016, \aap, 591, L3
      \item Manara, C.~F., Morbidelli, A., \& Guillot, T.\ 2018, \aap, 618, L3
      \item Masset, F.~S., D'Angelo, G., \& Kley, W.\ 2006a, \apj, 652, 730
        \item Masset, F.~S., Morbidelli, A., Crida, A., et al.\ 2006b, \apj, 642, 478
\item Morbidelli, A., Bottke, W.~F., Nesvorn{\'y}, D., et al.\ 2009, \icarus, 204, 558
\item Morbidelli, A., Lambrechts, M., Jacobson, S., et al.\ 2015, \icarus, 258, 418
\item Morbidelli, A., \& Nesvorn{\'y}, D.\ 2020, The Trans-neptunian Solar System, 25
  \item Najita, J.~R., \& Bergin, E.~A.\ 2018, \apj, 864, 168
\item Ormel, C.~W., \& Klahr, H.~H.\ 2010, \aap, 520, A43
\item Ormel, C.~W.\ 2017, Astrophysics and Space Science Library, 197
\item Paardekooper, S.-J., Baruteau, C., Crida, A., et al.\ 2010, \mnras, 401, 1950
  \item Paardekooper, S.-J., Baruteau, C., \& Kley, W.\ 2011, \mnras, 410, 293
\item Pinilla, P., Birnstiel, T., Ricci, L., et al.\ 2012, \aap, 538, A114
\item Pinte, C., Dent, W.~R.~F., M{\'e}nard, F., et al.\ 2016, \apj, 816, 25
  \item Pinte, C., Price, D.~J., M{\'e}nard, F., et al.\ 2020, \apjl, 890, L9
\item Riols, A., \& Lesur, G.\ 2019, \aap, 625, A108
\item Rosotti, G.~P., Tazzari, M., Booth, R.~A., et al.\ 2019, \mnras, 486, 4829
\item Sato, T., Okuzumi, S., \& Ida, S.\ 2016, \aap, 589, A15
  \item Shakura, N.~I., \& Sunyaev, R.~A.\ 1973, \aap, 500, 33
  \item Simon, J.~B., Armitage, P.~J., Youdin, A.~N., et al.\ 2017, \apjl, 847, L12
    \item Stevenson, D.~J.\ 1985, \icarus, 62, 4
    \item Tominaga, R.~T., Takahashi, S.~Z., \& Inutsuka, S.-. ichiro .\ 2019, \apj, 881, 53
    \item Trapman, L., Facchini, S., Hogerheijde, M.~R., et al.\ 2019, \aap, 629, A79
    \item Yang, C.-C., \& Zhu, Z.\ 2020, \mnras, 491, 4702
\item Youdin, A.~N., \& Lithwick, Y.\ 2007, \icarus, 192, 588
\item Wafflard-Fernandez, G., \& Baruteau, C.\ 2020, \mnras, doi:10.1093/mnras/staa379
  \item Wahl, S.~M., Hubbard, W.~B., Militzer, B., et al.\ 2017, \grl, 44, 4649
      \item Weber, P., P{\'e}rez, S., Ben{\'\i}tez-Llambay, P., et al.\ 2019, \apj, 884, 178
    \item Zhang, S., Zhu, Z., Huang, J., et al.\ 2018, \apjl, 869, L47

\end{itemize}

\end{document}